\documentclass[12pt,superscriptaddress]{article}
\usepackage{palatino,cite,epsfig,amsmath,amssymb}

\allowdisplaybreaks

\def\lsim{\raise0.3ex\hbox{$<$\kern-0.75em\raise-1.1ex\hbox{$\sim$}}}
\def\gsim{\raise0.3ex\hbox{$>$\kern-0.75em\raise-1.1ex\hbox{$\sim$}}}
\def\anti{\overline}
\def\beq{\begin{equation}}
\def\eeq{\end{equation}}
\def\bea{\begin{eqnarray}}
\def\eea{\end{eqnarray}}
\def\gam{\gamma}
\def\anti{\overline}
 
\def\br{BR}

\def\what{\widehat}
\def\hpm{H^{\pm}}
\def\ie{{\it i.e.}}
\def\gev{~{\rm GeV}}
\def\noi{\noindent}

\begin{document}
\vskip 1. truecm
\begin{flushright}
\noi UCD-03-4 \par
\noi IFIC/03-19 \par 
\noi LPT 03-35\par
\noi SHEP-03-08 \par 
\noi hep-ph/0305109 \par
\noi April, 2003 \par
\end{flushright}

\vspace{1cm}

\begin{center}

{\Large\bf Towards a No-Lose Theorem for NMSSM Higgs Discovery at the LHC}
 
\vspace{1cm}

{\sc Ulrich Ellwanger$^1$, John F. Gunion$^2$, Cyril Hugonie $^3$ and Stefano
Moretti$^4$}

\vspace*{1cm}

{\sl $^1$Laboratoire de Physique Th\'eorique, \\
Universit\'e de Paris XI, B\^atiment 210, F-91405 ORSAY Cedex, France
\vspace*{0.4cm}

$^2$Davis Institute for High Energy Physics, \\
University of California, Davis, California 95616, USA
\vspace*{0.4cm}

$^3$ AHEP Group, Instituto de F\'\i sica Corpuscular -- CSIC/Universitat de
Val\`encia, \\
Edificio Institutos de Investigaci\'on, Apartado de Correos 22085, E-46071
Val\`encia, Spain
\vspace*{0.4cm}

$^4$ Department of Physics and Astronomy, University of Southampton,\\ 
Southampton, SO17 1BJ, UK}

\end{center}

\vspace*{1cm}

\begin{abstract}
We scan the parameter space of the NMSSM for the observability of a Higgs boson
at the LHC with $300~{\rm fb}^{-1}$ integrated luminosity per detector, taking
the present LEP constraints into account. We focus on the
regions of parameter space for which 
none of the usually considered LHC detection modes
are viable due to the fact that the only light non-singlet (and, therefore,
potentially visible) Higgs boson, $h$, 
decays mainly to two CP-odd light Higgs bosons, $h\to a a$. We simulate the 
$WW\to h \to aa$ detection mode. We find
that this signal may be detectable at the LHC 
as a signal/background $\sim (500-1000)/300$ 
bump in the tail of a rapidly falling mass distribution. If further study
gives us confidence that the shape of the background
tail is predictable, then we can conclude that NMSSM Higgs detection at the LHC
will be possible throughout all of parameter space
by combining this signal with the  usual detection
modes previously simulated by ATLAS and CMS.
We also show that this $WW\to h\to aa$ signal
will be highly visible at the LC
due to its cleaner environment and high luminosity. We present a 
study of the production modes
and decay channels of interest at the LC.
\end{abstract}

\section{Introduction}\label{sec:Intro}

One of the most attractive supersymmetric models is the
Next to Minimal Supersymmetric Standard 
Model (NMSSM) \cite{Ellis:1988er,model,8r,9r},
which extends the MSSM by the introduction of just one single superfield,
$\what S$. When the scalar component of $\what S$
acquires a TeV scale vacuum expectation value (a very natural result in
the context of the model), 
the superpotential term $\what S \what H_u \what H_d$
generates an effective $\mu\what H_u \what H_d$
interaction for the Higgs doublet superfields. 
Such a term is essential for acceptable 
phenomenology. No other SUSY model generates this crucial 
component of the superpotential
in as natural a fashion. Thus, the phenomenological implications
of the NMSSM at future accelerators should be considered very seriously.
One aspect of this is the fact that the $h,H,A,\hpm$ Higgs sector
of the MSSM is extended so that there are three CP-even Higgs bosons
($h_{1,2,3}$, $m_{h_1}<m_{h_2}<m_{h_3}$), two CP-odd Higgs bosons ($a_{1,2}$,
$m_{a_1}<m_{a_2}$) (we assume that
CP is not violated in the Higgs sector)
and a charged Higgs pair ($h^\pm$). An important question
is then the extent to which the no-lose theorem for MSSM Higgs
boson discovery at the LHC (after LEP constraints) 
is retained when going to the NMSSM; \ie\
is the LHC guaranteed to find at least one of the 
$h_{1,2,3}$, $a_{1,2}$, $h^\pm$? 

One of the key ingredients in the no-lose theorem for MSSM Higgs
boson discovery is the fact that decays of the SM-like Higgs boson
to $AA$ are only possible if $m_A$ is quite small.
This is due to the fact that the masses of the $h,H,A,\hpm$
of the MSSM are closely tied to one another.
In particular, $m_A$ can only be small if $m_h$
is also relatively small 
and $m_{\hpm}\sim m_W$. In this region, it is the $H$
that is SM-like. The small-$m_A$ region is either excluded by
LEP because of non-observation of $hA$ production, or will
be detectable at the LHC via observation of $t\to b\hpm$ decays
in $t\anti t$ production. Most of the still unconstrained portion
of MSSM parameter space corresponds to the decoupling
region of $m_A\sim m_H\sim m_{\hpm}> 120-130\gev$ in
which the $h$ is SM-like and $m_h\lsim 120-130\gev$ so that $h\to AA$
decays are forbidden.  In the NMSSM, these strong mass
relations are weakened.  The $h_1$ or $h_2$ can be SM-like without
the $a_1$ necessarily being heavy.  As a result, $h_1\to a_1a_1$
or $h_2\to a_1a_1$ decays can be prominent and fall outside
the scope of the usual detection modes for the SM-like MSSM $h$
on which the MSSM no-lose LHC theorem largely relies.

In Ref.~\cite{Ellwanger:2001iw}, a partial no-lose
theorem for NMSSM Higgs boson discovery at the LHC was established.  
In particular, it was shown that the LHC would be able to detect at
least one of the NMSSM Higgs bosons (typically, one of the lighter
CP-even Higgs states) throughout the full parameter space of the model, 
excluding only those
parameter choices for which there is sensitivity to the
model-dependent decays of Higgs bosons to other Higgs bosons and/or
superparticles.  Here, we will retain the assumption
of a heavy superparticle spectrum and address the question of 
whether or not this
no-lose theorem can be extended to those regions of NMSSM parameter
space for which Higgs bosons can decay to other Higgs bosons.
We find that the parameter choices such
that the ``standard'' discovery modes fail
{\it would} allow Higgs boson discovery if
detection of $h\to aa$ decays is possible. (When
used generically, the symbol $h$ will now refer to
$h=h_1$, $h_2$ or $h_3$
and the symbol $a$ will refer to $a=a_1$ or $a_2$).  
Detection of $h\to aa$ will be difficult since each $a$ will decay
primarily to either $b\anti b$ (or 2 jets if $m_a<2m_b$) 
and $\tau^+\tau^-$, yielding final
states that will typically have large backgrounds at the LHC. Further,
a light $a$ can only be detected on its own at the LHC if it has a highly
enhanced Yukawa coupling to $b$ (or $t$) quarks.

Let us begin by reviewing our earlier procedure. For each
point in parameter space not ruled out by constraints from
LEP on Higgs boson production, $e^+ e^- \to Z h$ \cite{LEP} or
associated production $e^+ e^- \to h a$ \cite{LEPHA}, 
we calculate the Higgs boson masses and decay branching ratios including all
the relevant radiative corrections.  We eliminate parameter
choices for which one Higgs boson can decay to two other Higgs bosons
or a vector boson plus a Higgs boson. For the remaining regions
of parameter space, we then estimate the
statistical significances (computed as $N_{SD}=S/\sqrt B$ for a given mode) for
all Higgs boson detection modes so far studied at the LHC
\cite{CMS,ATLAS,6.2r,Zeppenfeld:2002ng}. These are (with $\ell=e,\mu$)

1) $g g \to h/a \to \gamma \gamma$;\par
2) associated $W h/a$ or $t \bar{t} h/a$ production with 
$\gamma \gamma\ell^{\pm}$ in the final state;\par
3) associated $t \bar{t} h/a$ production with $h/a \to b \bar{b}$;\par
4) associated $b \bar{b} h/a$ production with $h/a \to \tau^+\tau^-$;\par
5) $g g \to h \to Z Z^{(*)} \to$ 4 leptons;\par
6) $g g \to h \to W W^{(*)} \to \ell^+ \ell^- \nu \bar{\nu}$;\par
7) $W W \to h \to \tau^+ \tau^-$;\par
8) $W W \to h\to W W^{(*)}$.\par

\noindent
The outcome is that, for an integrated luminosity of $300~{\rm fb}^{-1}$ at
the LHC, there are still regions in the parameter space with $< 5\sigma$
expected statistical significance for all Higgs boson 
detection modes so far studied
in detail by ATLAS and CMS, {\it i.e.} modes 1) -- 6) but not the $W$-fusion
modes 7), 8). On the other hand, the expected statistical significance for at
least one of the non-$W$-fusion modes is always above $3.6\sigma$ at $300~{\rm
fb}^{-1}$, and the statistical significance obtained by combining (using the
naive Gaussian procedure) all the non-$W$-fusion modes is at least
$4.8\sigma$. However, we found that all such cases are quite observable (at
$\geq 10.1\sigma$) in one of the $W$-fusion modes (using theoretically
estimated statistical significances for these modes). For all points in the
scanned parameter space, statistical significances obtained by combining all
modes, including $W$-fusion modes, are always $\gsim 10.7\sigma$. Thus, NMSSM
Higgs boson 
discovery by just one detector with $L=300~{\rm fb}^{-1}$ is essentially
guaranteed for those portions of parameter space for which Higgs boson
decays to
other Higgs bosons or supersymmetric particles are kinematically forbidden.

In this work, we investigate the complementary part of the parameter
space, where {\it at least one} Higgs boson decays to other Higgs bosons. To be
more precise, we require at least one of the following decay modes to be
kinematically allowed:
\begin{eqnarray}
& i) \ h \to h' h' \; , \quad ii) \ h \to a a \; , \quad iii) \ h \to h^\pm
h^\mp \; , \quad iv) \ h \to a Z \; , \nonumber \\
& v) \ h \to h^\pm W^\mp \; , \quad vi) \ a' \to h a \; , \quad vii) \ a \to h
Z \; , \quad viii) \ a \to h^\pm W^\mp \; .
\end{eqnarray}
The observability at the LHC of Higgs bosons that decay in these ways
has not been studied, except for some
particular MSSM cases that are not applicable for the masses and Yukawa
coupling strengths that are most relevant in the NMSSM context.
After searching those regions of parameter space for which one or
more of the decays $ i) - viii)$ is allowed, we found that the
only subregions that would not be excluded by modes 1) -- 8)
correspond to NMSSM parameter
choices for which (a) there is a light CP-even Higgs boson
with substantial doublet content that decays mainly to two
still lighter CP-odd Higgs states, $h\to aa$, and (b) all the other
Higgs states are either dominantly singlet-like,
implying highly suppressed production rates,
or relatively heavy, decaying to $t\anti t$ or to 
one of the ``difficult'' modes
$i) - viii)$.  In such cases, it seems evident that the best
opportunity for detecting at least one of the NMSSM Higgs
bosons is to employ $WW\to h$ production and develop techniques
for extracting a visible signal for the $h\to aa$ final state.
We have performed a detailed simulation of this signal and
find that its detection may be possible
after accumulating $300~{\rm fb}^{-1}$ in both the ATLAS and CMS
detectors.  The nominal statistical significance achieved
is $>30\sigma$ with $S/B > 1$. However,
the signal only emerges on the tail
of a rapidly falling background. A thorough understanding of
and confidence in the shape of this background tail would
be needed to claim detection of the $h\to aa$ signal.
If further study and, ultimately, 
actual measurements of the background outside the signal region
result in such confidence then the combined search for
this signal and the signals for Higgs bosons in 
the previously studied modes 1) -- 8) would result in detection
of at least one of the NMSSM Higgs bosons throughout the entire
NMSSM parameter space (assuming no large branching ratios
for Higgs boson decays to sparticles).
Having reached this conclusion, we turn to a study of the role
that a future LC will play in exploring the NMSSM Higgs sector
for those parameter choices for which 
detecting the $h\to aa$ signal is so critical. 
We find, regardless of what happens at the LHC,
that the LC is guaranteed to find one or more Higgs
bosons, including detection of the $h\to aa$ signal
in several different production channels, leading to
the possibility of measuring the $h\to aa$ branching ratio.

In section~\ref{sec:Param}, we review the basic set up of the NMSSM
as well as our scanning procedure for the NMSSM parameter space.
In section~\ref{sec:InvPts}, we give a selection of benchmark points for which
all statistical significances at the LHC are below $5\sigma$
after employing the standard discovery modes 1) -- 8). We then
discuss possible new discovery
channels in section~\ref{sec:NewChan} together with our simulation
results both at the LHC and the LC. Conclusions are given in section
\ref{sec:Conc}.

\section{NMSSM Parameters}\label{sec:Param}

We consider the simplest version of the NMSSM \cite{Ellis:1988er,8r,9r}, where the term
$\mu \widehat H_1 \widehat H_2$ in the superpotential of the MSSM is replaced
by (we use the notation $\widehat A$ for the superfield and $A$ for its scalar
component field)

\begin{equation}\label{2.1r}
\lambda \widehat H_1 \widehat H_2 \widehat S\ + \ \frac{\kappa}{3} \widehat S^3
\ \ ,
\end{equation}

\noindent so that the superpotential is scale invariant. We make no assumption
on ``universal'' soft terms. Hence, the five soft supersymmetry breaking terms

\begin{equation}\label{2.2r}
m_{H_1}^2 H_1^2\ +\ m_{H_2}^2 H_2^2\ +\ m_S^2 S^2\ +\ \lambda
A_{\lambda}H_1 H_2 S\ +\ \frac{\kappa}{3} A_{\kappa}S^3
\end{equation}

\noindent are considered as independent. The masses and/or couplings of
sparticles are assumed to be such that their contributions to the loop diagrams
inducing Higgs boson 
production by gluon fusion and Higgs boson decay into $\gamma \gamma$
are negligible. In the stop sector, which appears in the radiative corrections
to the Higgs potential, we chose the soft masses $m_Q = m_T \equiv M_{susy}= 1$
TeV, and vary the stop mixing parameter 

\begin{equation}\label{2.4r}
X_t \equiv 2 \ \frac{A_t^2}{M_{susy}^2+m_t^2} \left ( 1 -
\frac{A_t^2}{12(M_{susy}^2+m_t^2)} \right ) \ .
\end{equation} 

\noindent As in the MSSM, the value $X_t = \sqrt{6}$ -- so called maximal
mixing -- maximizes the radiative corrections to the Higgs boson
masses, and we found
that it leads to the most challenging points in the parameter space of the
NMSSM. 

Assuming that the Higgs sector is CP conserving, the independent parameters of 
the model are thus: $\lambda, \kappa, m_{H_1}^2, m_{H_2}^2, m_S^2, A_{\lambda}$
and $A_{\kappa}$. For purposes of scanning and analysis, it is more convenient
to eliminate $m_{H_1}^2$, $m_{H_2}^2$ and $m_S^2$ in favor of $M_Z$,
$\tan\beta$ and $\mu_{\rm eff} = \lambda \langle S \rangle$ through the three
minimization equations of the Higgs potential (including the dominant 1- and
2-loop corrections \cite{Ellwanger:1999ji}) and to scan over the six
independent parameters 

\begin{equation}\label{2.5r}
\lambda, \kappa, \tan\beta, \mu_{\rm eff}, A_{\lambda}, A_{\kappa}\ .
\end{equation}

We adopt the convention $\lambda,\kappa > 0$, in which $\tan\beta$ can have
either sign. The absence of Landau singularities for $\lambda$ and $\kappa$
below the GUT scale ($\sim 2\times10^{16}$~GeV) imposes upper bounds on these
couplings at the weak scale, which depend on the value of $h_t$ and hence of
$\tan\beta$ \cite{Ellis:1988er,8r}. Using $m_t^{\rm pole} = 175$~GeV, one finds
$\lambda_{\rm max} \sim 0.69$ and $\kappa_{\rm max} \sim 0.62$ for intermediate
values of $\tan\beta$. In addition, we require $|\mu_{\rm eff}|\ >\ 100$~GeV;
otherwise a light chargino would have been detected at LEP. (The precise lower
bound on $|\mu_{\rm eff}|$ depends somewhat on $\tan\beta$ and the exact
experimental lower bound on the chargino mass; however, our subsequent results
do not depend on the precise choice of the lower bound on $|\mu_{\rm eff}|$.)

We have performed a numerical scanning over the free parameters, which were
randomly chosen in the following intervals:

\begin{eqnarray} \label{param}
& & 0.1 < \lambda < \lambda_{\rm max} \; , \quad 0.001 < \kappa < \kappa_{\rm
max} \; , \quad 2.5 < |\tan\beta| < 10 \; , \nonumber \\
& & 100~{\rm GeV} < |\mu_{\rm eff}| < 500~{\rm GeV} \; , \; 0 < |A_\lambda| <
500~{\rm GeV} \; , \; 0 < |A_\kappa| < 500~{\rm GeV} \; .
\end{eqnarray}

\noindent For each point, we computed the CP-even and CP-odd Higgs boson masses
and mixings, taking into account radiative corrections up to the dominant two
loop terms, as described in \cite{Ellwanger:1999ji}. The five mass eigenstates
are denoted by $h_i$ ($i=1,2,3$) and $a_j$ ($j=1,2$) for CP-even and CP-odd
eigenstates respectively, with masses $m_{h_i}$ and $m_{a_j}$ in increasing
order. We have eliminated parameter choices
that lead to violation of the  
LEP constraints on Higgs boson 
production $e^+ e^- \to Z h_i$ ($i=1$, $2$ or $3$)
\cite{LEP}. These constraints give an upper bound on the $ZZh_i$
reduced coupling, $R_i$, as a function of $m_{h_i}$. (The reduced coupling
$R_i$ is defined as the coupling $ZZh_i$ divided by the corresponding Standard
Model coupling and is the equivalent of $\sin(\beta-\alpha)$ in the MSSM.) We
have also kept only parameters leading to
consistency with the exclusion limit from LEP on Higgs bosons associated
production $e^+ e^- \to h_i a_j$ \cite{LEPHA}. This provides an upper bound on
the $Zh_ia_j$ reduced coupling, $R'_{ij}$, as a function of $m_{h_i}+m_{a_j}$
for $m_{h_i} \simeq m_{a_j}$ \cite{LEPHA}. (The $Zh_ia_j$ reduced coupling is
the equivalent of $\cos(\beta-\alpha)$ in the MSSM.) Finally, we calculated the
charged Higgs boson 
mass $m_{h^\pm}$ and required $m_{h^\pm} > 155$~GeV, so that $t
\to h^\pm b$ would not be seen.

In order to probe the complementary part of the parameter space as compared to
the scanning of Ref. \cite{Ellwanger:2001iw}, we required that at least one of
the ``difficult'' decay modes $i) - viii)$ displayed in the previous section is
allowed. For each Higgs state, we have calculated the branching ratios in the
usual decay modes as well as in the difficult modes $i) - viii)$ when
kinematically allowed. In doing so, we have included all relevant higher-order
QCD corrections \cite{Djouadi:1995gt} using an adapted version of the FORTRAN
code HDECAY \cite{Djouadi:1997yw}. We then estimated the expected statistical
significances at the LHC in all Higgs boson detection modes 1) -- 8) as described in
the next section.

\section{Invisible Points at the LHC}\label{sec:InvPts}

From the known couplings of the NMSSM Higgs scalars to gauge bosons and
fermions it is straightforward to compute their production rates in gluon-gluon
fusion and various associated production processes, as well as their partial
widths into $\gamma \gamma$, gauge bosons and fermions, either relative to a SM
Higgs scalar or relative to the MSSM $h, H$ and/or $A$. This allows us to apply
``NMSSM corrections'' to the processes 1) -- 8) of section~\ref{sec:Intro}.
These NMSSM corrections are computed in terms of the following ratios: For the
scalar Higgs bosons, $R_i$ is the ratio of the coupling of the $h_i$ to vector
bosons as compared to that of a SM Higgs boson, and $t_i$, $b_i$ are the
corresponding ratios of the couplings to top and bottom quarks. 
(The coupling ratio for $\tau$ leptons is always the same as for $b$ quarks,
\ie\ 
$\tau_i=b_i$). Note that we always have $|R_i| < 1$, but $t_i$ and $b_i$ can be
larger, smaller or even differ in sign with respect to the SM. For the CP-odd
Higgs bosons, there is no tree-level coupling to the $VV$ states; $t'_j$ and
$b'_j$ are defined as the ratio of the $i\gamma_5$ couplings for $t\bar{t}$ and
$b\bar{b}$, respectively, relative to SM-like strength. We have also taken into
account the contributions of the ``difficult'' decay modes $i) - viii)$ to the
total decay width when kinematically allowed.

The expected statistical significances for the processes 1) $g g \to h/a \to
\gamma \gamma$ and 6) $g g \to h \to W W^{(*)} \to \ell^+ \ell^- \nu \bar{\nu}$ are
computed from results for the SM Higgs boson taken from Ref.~\cite{CMS},
Fig.~1. The application of the NMSSM corrections using $R_i$, $t_i$ and $b_i$
(which determine $\Gamma(g g \to h_i)$, $BR(h_i \to \gamma \gamma)$ and $BR(h_i
\to W W^*)$) is straightforward in these two cases. The expected statistical
significances for process 2) $W h/a$ or $t \bar{t} h/a$ with 
$\gamma \gamma\ell^{\pm}$ in the final state, 
are taken from the same figure. One finds that
$W h_i$ and $t \bar{t}h_i$ production contribute with roughly equal weight to
the SM signal. This allows us to decompose the expected significance into the
corresponding production processes, apply the NMSSM corrections, and then
recombine the production processes.

For the Standard Model Higgs boson, 
the expected statistical significances for process 3)  $t
\bar{t} h/a$ with $h/a \to b \bar{b}$, are taken from table 19-8 in
Ref.~\cite{ATLAS}, with the extension to Higgs boson
masses above 120~GeV as provided
by \cite{Sapinski}. For the standard model process 5) $g g \to h \to Z Z^{(*)}
\to$ 4 leptons, we again use Ref.~\cite{ATLAS}, tables 19-18 and 19-21. In both
cases, the application of the NMSSM corrections is straightforward. The
estimation of the statistical significances for the process 4) $b \bar{b} h/a$
with $h/a \to \tau^+ \tau^-$, is more involved. Figure 19-62 of
Ref.~\cite{ATLAS} give the $5 \sigma$ contours in the $\tan\beta - m_A$ plane
of the MSSM. The critical issue is how much of these $5\sigma$ signals derive
from $g g \to H + g g \to A$ production and how much from associated $b \bar{b}
H + b \bar{b} A$ production, and how each of the $g g$ fusion and $b \bar{b}$
associated production processes are divided up between $H$ and $A$. For more
details on our procedure, see Ref.~\cite{Ellwanger:2001iw}.

Results for the statistical significances of the $h_i$ signals in modes 7) $W W
\to h \to \tau^+\tau^-$ and 8) $W W \to h\to W W^{(*)}$ were similarly
obtained by rescaling the theoretical results of \cite{Zeppenfeld:2002ng}. 

In the case of Higgs states close in mass, the individual statistical
significances have to be recombined. We used the procedure of
Ref.~\cite{Richter-Was:1996ak}, which gives the combined statistical
significance $N_{SD,12}$ as a function of the individual statistical
significances $N_{SD,1}, N_{SD,2}$ for two states with masses $m_1, m_2$ if
$|m_1 - m_2|\ < \ \Delta_k \frac{m_1 + m_2}{2}$, where $\Delta_k$ is the mass
resolution of the decay mode $k)$:

\begin{equation}
N_{SD,12}^2 = N_{SD,1}^2 + N_{SD,2}^2 + 2 \frac {N_{SD,1} N_{SD,2}} {1 +
(\frac{2}{\Delta_k} \frac{m_1 - m_2}{m_1 + m_2})^2} \; .
\end{equation}
For the processes 1) -- 8), we assumed $\Delta_{1,2,5,6} = 1 \%, 
\Delta_{3,7,8} = 10 \%,
$ and $\Delta_4 = 15\%
$.

\begin{table}[p]
\begin{center}
\footnotesize
\begin{tabular} {|l|l|l|l|l|l|l|} 
\hline
Point Number & 1 & 2 & 3 & 4 & 5 & 6  \\
\hline \hline
Bare Parameters &\multicolumn{6}{c|}{} \\
\hline
$\lambda$            & 0.2872 & 0.2124 & 0.3373 & 0.3340 & 0.4744 & 0.5212 \\
\hline
$\kappa$             & 0.5332 & 0.5647 & 0.5204 & 0.0574 & 0.0844 & 0.0010 \\
\hline
$\tan\beta$          &   2.5  &   3.5  &   5.5  &    2.5 &    2.5 & 2.5 \\
\hline
$\mu_{\rm eff}~({\rm GeV})$&    200 &    200 &    200 &    200 &    200 & 200 \\
\hline
$A_{\lambda}~({\rm GeV})$  &    100 &      0 &     50 &    500 &    500 & 500 \\
\hline
$A_{\kappa}~({\rm GeV})$   &      0 &      0 &      0 &      0 &      0 & 0 \\
\hline \hline
CP-even Higgs Boson Masses and Couplings &\multicolumn{6}{c|}{} \\
\hline \hline
$m_{h_1}$~(GeV)      &    115 &    119 &    123 &     76 &     85 &  51\\
\hline
$R_1 $               &   1.00 &   1.00 &  -1.00 &   0.08 &   0.10 &  -0.25\\
\hline
$t_1 $               &   0.99 &   1.00 &  -1.00 &   0.05 &   0.06 &  -0.29\\
\hline
$b_1 $               &   1.06 &   1.05 &  -1.03 &   0.27 &   0.37 &  0.01\\
\hline
Relative 
gg Production Rate   &   0.97 &   0.99 &   0.99 &   0.00 &   0.01 &  0.08\\
\hline
$\br(h_1\to 
b\anti b)$           &   0.02 &   0.01 &   0.01 &   0.91 &   0.91 &  0.00\\
\hline
$\br(h_1\to 
\tau^+\tau^-)$      &   0.00 &   0.00 &   0.00 &   0.08 &   0.08 &  0.00\\
\hline
$\br(h_1\to a_1 a_1)$&   0.98 &   0.99 &   0.98 &   0.00 &   0.00 &  1.00\\
\hline \hline

$m_{h_2}$~(GeV)      &    516 &    626 &    594 &    118 &    124 &  130\\
\hline
$R_2 $               &  -0.03 &  -0.01 &   0.01 &  -1.00 &  -0.99 &  -0.97\\
\hline
$t_2 $               &  -0.43 &  -0.30 &  -0.10 &  -0.99 &  -0.99 &  -0.95\\
\hline
$b_2 $               &   2.46 &  -3.48 &   3.44 &  -1.03 &  -1.00 &  -1.07\\
\hline
Relative
gg Production Rate   &   0.18 &   0.09 &   0.01 &   0.98 &   0.99 &  0.90\\
\hline
$\br(h_2\to 
b\anti b)$           &   0.01 &   0.04 &   0.04 &   0.02 &   0.01 &  0.00\\
\hline
$\br(h_2\to 
\tau^+\tau^-)$      &   0.00 &   0.01 &   0.00 &   0.00 &   0.00 &  0.00\\
\hline
$\br(h_2\to a_1 a_1)$&   0.04 &   0.02 &   0.83 &   0.97 &   0.98 &  0.96\\
%\hline
%$\br(h_2\to h_1 h_1)$&   0.01 &   0.01 &   0.00 &   0.00 &   0.00 &  0.04\\
\hline \hline

$m_{h_3}$~(GeV)      &    745 &   1064 &    653 &    553 &    554 &  535\\
\hline \hline

CP-odd Higgs Boson Masses and Couplings &\multicolumn{6}{c|}{} \\
%and Couplings &\multicolumn{6}{c|}{} \\
\hline \hline
$m_{a_1}$~(GeV)      &     56 &      7 &     35 &     41 &     59 &  7\\
\hline
$t_1' $               &   0.05 &   0.03 &   0.01 &  -0.03 &  -0.05 &  -0.06\\
\hline
$b_1' $               &   0.29 &   0.34 &   0.44 &  -0.20 &  -0.29 &  -0.39\\
\hline
Relative
gg Production Rate   &   0.01 &   0.03 &   0.05 &   0.01 &   0.01 &  0.05\\
\hline
$\br(a_1\to 
b\anti b)$           &   0.92 &   0.00 &   0.93 &   0.92 &   0.92 &  0.00\\
\hline
$\br(a_1\to 
\tau^+\tau^-)$      &   0.08 &   0.94 &   0.07 &   0.07 &   0.08 &  0.90\\
\hline \hline

$m_{a_2}$~(GeV)      &    528 &    639 &    643 &    560 &    563 &  547\\
\hline 
Charged Higgs  
Mass (GeV)           &    528 &    640 &    643 &    561 &    559 &  539\\
\hline\hline
Most Visible Process No. &  2 ($h_1$) &  2 ($h_1$) &  8
                  ($h_1$) &  2 ($h_2$) &  8 ($h_2$)  &  8 ($h_2$)\\
\hline            
Significance at 300~${\rm fb}^{-1}$
                     &   0.48 &   0.26 &   0.55 &   0.62 &  0.53  & 0.16\\
\hline
\end{tabular}
\end{center}
\caption{\label{tpoints}\footnotesize
Properties of selected scenarios that could escape detection
at the LHC. In the table, $R_i$, $t_i$ and $b_i$ are
the ratios of the $h_i$ couplings to $VV$, $t\anti t$ and $b\anti b$,
respectively, as compared to those of a SM Higgs boson
with the same mass; $t_1'$ and $b_1'$
denote the magnitude of the $i\gam_5$ couplings of $a_1$ 
to $t\anti t$ and $b\anti b$ normalized
relative to the magnitude  of the $t\anti t$ and $b\anti b$ SM Higgs couplings.
We also give the production
for $gg\to h_i$ fusion
relative to the $gg$ fusion rate for
a SM Higgs boson with the same mass. Important absolute branching
ratios are displayed. For points 2 and 6, $\br(a_1\to jj)\simeq 
1-\br(a_1\to \tau^+\tau^-)$.
For the heavy $h_3$ and $a_2$, we give only their masses.
In the case of the points 2 and 6, decays of $a_1$
into light quarks start to contribute. For all points 1 -- 6, the statistical
significances for the detection of any Higgs boson in any of the channels 1) --
8) (as listed in the introduction) are tiny; their maximum is indicated in the
last row, together with the process number and the corresponding Higgs state.}
\end{table}

Using the above procedures, for each point in the parameter space of the NMSSM
we obtain the statistical significances predicted for an integrated luminosity
of $100~{\rm fb}^{-1}$ for each of the detection modes 1) -- 8). In order to
obtain the statistical significances for the various detection modes at 
$300~{\rm fb}^{-1}$, 
we multiply the $100~{\rm fb}^{-1}$ statistical significances
by $\sqrt{3}$ in the cases 1), 2), 3), 5) and 6), but only by a factor of $1.3$
in the cases 4), 7) and 8). That such a factor is appropriate for mode 4), see,
for example, Fig.~19-62 in \cite{ATLAS}. Use of this same factor for modes 7)
and 8) is simply a conservative guess.

In our set of randomly scanned points defined in
section~\ref{sec:Param}, we selected those for which all the
statistical significances [including $WW$-fusion modes 7) and 8)] are below
$5\sigma$. We obtained a lot of points, all with similar
characteristics. Namely, in the Higgs spectrum, we always have a very
SM-like CP-even Higgs boson with a mass between 115 and 135~GeV ({\it
  i.e.} above the LEP limit), which can be either $h_1$ or $h_2$, with
a reduced coupling to the gauge bosons $R_1 \simeq 1$ or $R_2\simeq 1$,
respectively. This state
decays dominantly to a pair of (very) light CP-odd states, $a_1a_1$,
with $m_{a_1}$ between 5 and 65~GeV\footnote{In some 
rare cases, one has $h_2$ decaying mainly to $h_1h_1$, but then the
LHC significances corresponding to the discovery channels 1) --- 8) 
are quite close to 5, so we have not considered cases of this type.}.  
The singlet component of $a_1$
has to be small in order to have a large $h_1 \to a_1 a_1$ or $h_2\to a_1a_1$
branching ratio when the
$h_1$ or $h_2$, respectively, is the SM-like Higgs boson. 
Further, when the $h_1$ or $h_2$ is very SM-like, one has
$R'_{11}\simeq 0$ or $R'_{21} \simeq 0$, respectively,
so that the $e^+ e^- \to h_1 a_1$ or $e^+e^-\to h_2 a_1$
associated production rate is very small and 
these processes thus place no constraint 
on the light CP-odd state at LEP. We have selected six
difficult benchmark points. For points 1 -- 3, $h_1$ is the SM-like
CP-even state, while for points 4 -- 6 it is $h_2$. We have chosen the
points so that $h_{1,2}$ and $a_1$ have different masses. The main
characteristics of the benchmark points are displayed in
table~\ref{tpoints}. Note the large branching ratios
of the SM-like Higgs boson ($h_1$ for points 1 -- 3 and $h_2$
for points 4 --6) to decay to $a_1a_1$. 
Branching ratios for
the non-SM-like $h_2$ (points 1 -- 3) or $h_1$ (points 4 -- 6)
are also shown.  For points 4 -- 6, with $m_{h_1}<100\gev$,
the $h_1$ is mainly singlet.  As a result, $R'_{11}$ is very
small and thus there are no LEP constraints on the $h_1$ and $a_1$
from $e^+e^-\to h_1 a_1$ production.

In the case of the points 1 -- 3, the tabulated 
branching ratios of the $h_2$ are all very small.
Even though the $h_2$ is heavy, it has very weak $WW,ZZ$ coupling
and $\br(h_2\to ZZ+WW)\sim 0.01$. Further, the $h_2$ production rate
due to $gg$ fusion is greatly suppressed, as indicated in the table,
as is its production rate from $WW,ZZ$ fusion. This means that
searches for the $h_2$ in the $ZZ\to 4\ell$ final state 
are impossible. 
In fact, the $h_2$ decays primarily to $t\anti t$ with the 2nd
most important mode being $Za_1$.   The only search channel
for the $h_2$ that might be worth further investigation appears
to be the $b\anti b h_2$ Yukawa ``radiation'' process with $h_2\to t\anti t$.
Even though the $b\anti b h_2$ Yukawa couplings are somewhat
enhanced, the overall production rate will not be that large and
we anticipate that the signal for the $h_2$ in this mode would
be quite marginal.  Regarding the LC, since $m_{h_2}\gsim 500\gev$,
$\sqrt{s_{e^+e^-}}$ substantially above the $500\gev$
level would be required. Even for such $\sqrt{s_{e^+e^-}}$, 
$h_2$ detection would
be very challenging since
the $ZZh_2,WWh_2$ couplings are very suppressed and the $b\anti b h_2$
Yukawa coupling is only somewhat enhanced.    
Thus, we have focused on the $a_1a_1$ decays
of the $h_1$  for points 1 -- 3.

For points 4 -- 6, even though the $h_1$ is quite light, it 
is dominantly singlet-like and therefore has
suppressed couplings to all relevant SM particles, $WW,ZZ$, $b\anti b$
and $t\anti t$.  As a result, it has very low production rates
and its decays (primarily to either $b\anti b$, points 4 and 5,
or $a_1a_1$, point 6) will not produce particularly distinctive
final states, especially given the smallness of $m_{h_1}$.  Thus,
we believe the $h_1$ would
be very difficult to detect at the LHC. 
At the LC, the $h_1$ has $ZZ$ coupling sufficient
for detection only in the case of point 6. 
Given the complicated nature of the $h_1\to a_1a_1$ final state in this
case (including the fact that the $a_1$ is so light that $a_1\to \tau^+\tau^-$
is dominant) detection would have to rely on the $ZX$ reconstructed
recoil mass technique. Thus, for points 4 -- 6,
we will focus on searching for the SM-like $h_2$
in its dominant $h_2\to a_1a_1$ decay mode. 

Of course, we should also consider whether or not the $h_3$ or $a_2$
might be detectable for points 1 -- 6. 
Since $m_{h_3}$ is very large in all cases, one would need
to rely on the $h_3\to ZZ \to 4\ell$ mode [mode 5) in our first list]
for its detection.  But,
the $gg$ and $WW$ fusion production rates are suppressed and
$\br(h_3\to ZZ)<0.01$.  Mode 5) for the $h_3$ contributes negligibly
to the net statistical significance in the last column 
of table~\ref{tpoints}. At the LC, $\sqrt{s_{e^+e^-}}$ substantially above
$500\gev$ would be required at all points.  Detection
of the $a_2$ would appear to be even more difficult. If a super
high energy LC is eventually built, the pair production modes
such as $e^+e^-\to Z\to h_3 a_2$ will be substantial and produce
very viable signals.

In the case of points 2 and 6, it should be noted that the $a_1\to \tau^+\tau^-$ decays are dominant, with $a_1\to jj$ decays making up most of the rest.
For points 1 and 3 -- 5 for which $\br(a_1\to b\anti b)$ is substantial,
the $b$ jets will not be that energetic and tagging will be somewhat inefficient. Thus, we have chosen not to implement $b$-tagging as part of 
the experimental procedures detailed in the next section.

Finally, we should consider whether or not a direct search for the $a_1$
might be feasible. For our points 1 -- 6, the $a_1$ has suppressed
couplings to both $b\anti b$ and $t\anti t$ and will thus be very
weakly produced via Yukawa radiation and would not be detectable
in this way at either the LHC or the LC~\cite{Grzadkowski:1999wj}.  
The only processes
that would have significant rate are those relying on
the $ZZa_1a_1$ and $WW a_1a_1$ quartic couplings required by gauge
invariance.  At the LHC, the rate for, say, $WW\to a_1a_1$
is not exactly large and, as we shall see,
there is a large background from $t\anti t$ production.
Rates for continuum $WW\to a_1a_1$ production
at the LC are substantial for a light $a_1$~\cite{Farris:2002ny}, 
but again we shall see that there
is a large background.  In the
next section, we demonstrate that 
the constraint that $M_{a_1a_1}\sim m_{h}$ with $m_{h}\sim 100\gev$
is absolutely critical in order to have any hope of extracting a signal
in $WW\to h$ production at both the LHC and the LC.

\section{New Channels at the LHC and LC}\label{sec:NewChan}
\def\cO#1{{\cal{O}}\left(#1\right)}
\def\nn {\nonumber}
\newcommand{\be}{\begin{equation}}
\newcommand{\ee}{\end{equation}}
\newcommand{\ba}{\begin{array}}
\newcommand{\ea}{\end{array}}
\newcommand{\bi}{\begin{itemize}}
\newcommand{\ei}{\end{itemize}}
\newcommand{\bn}{\begin{enumerate}}
\newcommand{\en}{\end{enumerate}}
\newcommand{\bc}{\begin{center}}
\newcommand{\ec}{\end{center}}
\newcommand{\ul}{\underline}
\newcommand{\ol}{\overline}
\newcommand{\ar}{\rightarrow}
\newcommand{\sm}{${\cal {SM}}$}
\newcommand{\as}{\alpha_s}
\newcommand{\aem}{\alpha_{em}}
\newcommand{\ycut}{y_{\mathrm{cut}}}
\newcommand{\susy}{{{SUSY}}}
\newcommand{\Dir}{\kern -6.4pt\Big{/}}
\newcommand{\Dirin}{\kern -10.4pt\Big{/}\kern 4.4pt}
\newcommand{\DDir}{\kern -10.6pt\Big{/}}
\newcommand{\DGir}{\kern -6.0pt\Big{/}}
\def\Ecm{\ifmmode{E_{\mathrm{cm}}}\else{$E_{\mathrm{cm}}$}\fi}
\def\gluino{\ifmmode{\mathaccent"7E g}\else{$\mathaccent"7E g$}\fi}
\def\photino{\ifmmode{\mathaccent"7E \gamma}\else{$\mathaccent"7E \gamma$}\fi}
\def\mgluino{\ifmmode{m_{\mathaccent"7E g}}
             \else{$m_{\mathaccent"7E g}$}\fi}
\def\taugluino{\ifmmode{\tau_{\mathaccent"7E g}}
             \else{$\tau_{\mathaccent"7E g}$}\fi}
\def\mphotino{\ifmmode{m_{\mathaccent"7E \gamma}}
             \else{$m_{\mathaccent"7E \gamma}$}\fi}
\def\ML{\ifmmode{{\mathaccent"7E M}_L}
             \else{${\mathaccent"7E M}_L$}\fi}
\def\MR{\ifmmode{{\mathaccent"7E M}_R}
             \else{${\mathaccent"7E M}_R$}\fi}

\def\lsim{\buildrel{\scriptscriptstyle <}\over{\scriptscriptstyle\sim}}
\def\gsim{\buildrel{\scriptscriptstyle >}\over{\scriptscriptstyle\sim}}
\def\Jnl #1#2#3#4 {#1 {\bf #2} (#3) #4}
\def\NPB {{\rm Nucl. Phys.} {\bf B}}
\def\PLB {{\rm Phys. Lett.}  {\bf B}}
\def\PRL {\rm Phys. Rev. Lett.}
\def\PRD {{\rm Phys. Rev.} {\bf D}}
\def\ZPC {{\rm Z. Phys.} {\bf C}}
\def\EPJC {{\rm Eur. Phys. J.} {\bf C}}
\def\Ord{\lower .7ex\hbox{$\;\stackrel{\textstyle <}{\sim}\;$}}
\def\OOrd{\lower .7ex\hbox{$\;\stackrel{\textstyle >}{\sim}\;$}}

As we have already stressed, for the 
points summarized in table~\ref{tpoints} the $a_1$ is light
and decays almost entirely 
into $b\anti b$ (or $jj$ for points 2 and 6) and $\tau^+\tau^-$.
The possible final states are thus $b\anti b b\anti b$ (or $4j$
for points 2 and 6), $b\anti b \tau^+\tau^-$ (or $2j\tau^+\tau^-$)
and $\tau^+\tau^-\tau^+\tau^-$.
A $4b$-signal would be burdened by a large QCD background
even after implementing $b$-tagging.  A $4j$-signal would be completely
swamped by QCD background.
Meanwhile,
the $4\tau$-channel would not allow one to reconstruct 
the $h_1,h_2$ resonances.
Hence, in attempting to cover these problematic points of the NMSSM parameter
space at the LHC, we will focus in this study on the $2b2\tau$ (or $2j2\tau$)
signature. The extra factor of two in 
$BR(a_1a_1\to b\anti b \tau^+\tau^-)=2BR(a_1\to b\anti b)BR(a_1\to \tau^+\tau^-)$ 
relative to 
$BR(a_1a_1\to b\anti b b\anti b)=[BR(a_1\to b\anti b)]^2$
and $BR(a_1a_1\to \tau^+\tau^-\tau^+\tau^-)=[BR(a_1\to\tau^+\tau^-)]^2$
 (and similarly with $b\anti b \to jj$
for points 2 and 6) is an additional advantage.
It compensates in part
(or further increases) the Yukawa suppression (or enhancement) 
with respect to the
$h_1/h_2\to a_1a_1\to 4b(4\tau)$ decay rate.\footnote{Recall that 
it is the running $b$-quark mass (evaluated at
the $a_1$ mass) that enters the Yukawa coupling, rather than the pole mass.}
In addition, we will be looking
at $\tau$'s decaying leptonically to electrons and muons, yielding some 
amount of missing (transverse) momentum, $p_{\rm miss}^T$, that could
be projected onto the visible $e,\mu$-momenta in an attempt to reconstruct
the parent $\tau$-direction. Since for points 2 and 6 the $a_1$
does not decay to $b\anti b$ and since the $b$ and $\bar b$ that do
come from $a_1$ are not very energetic given the modest $m_{a_1}$ mass
for points 1 and 3 -- 5, we will not employ $b$-tagging
as part of our analysis.

\subsection{Results for the LHC}

Detection of $WW^*$ and $\tau^+\tau^-$ 
decays of a relatively light Higgs boson
with SM-like $WW$ coupling is straightforward at the LHC 
in the vector-vector fusion 
production mode [\ie\, the Higgs production channel entering the signatures
8) and 9)] when the light Higgs boson has mass above the LEP limits
and decays directly to $\tau^+\tau^-$ 
(the status of the $b\anti b$ mode has not yet been determined).
Because the significance of this type of signal
for a SM-like Higgs boson is very large, we consider here the same $WW$-fusion
production mode in the context of the NMSSM. (We reemphasize that
the $h_1$ [cases 1 -- 3] or $h_2$ [cases 4 -- 6]
has nearly full SM strength coupling to $WW$.) 
However, the $b\anti b\tau^+\tau^-$ (or $2j\tau^+\tau^-$,
for points 2 and 6) final state of relevance 
is more complex and subject to larger backgrounds than is a simple 
$\tau^+\tau^-$ final state.  Further,  
the $a_1$ masses of interest are only half as large 
as the Higgs masses for which the direct $h \to \tau^+\tau^-$ signals
are viable.
In order to extract the $2j2\tau$ NMSSM Higgs boson
signature from the central detector region, we have exploited
forward and backward jet tagging
on the light quarks emerging after the
double $W$-strahlung preceding $WW$-fusion, 
the utility of which is reviewed in Ref.~\cite{jettagging}. 
If we require two additional forward/backward jets, 
it is clear that the leading background
is due to $t\anti t$ production (since we are assuming a heavy
SUSY spectrum) and decay  via the purely SM process,
\begin{equation}\label{background}
gg\to t\bar t\to b\bar b W^+W^-\to b\bar b \tau^+\tau^- + p_{\rm miss}^T,
\end{equation}
in association with forward and backward jet radiation. 

In summary, at the LHC, the signature is 
\begin{itemize}
\item 2 forward/backward jets, at least 2 central jets, $p^T_{\rm{miss}}$ and 
a $\tau^+\tau^-$ pair decaying leptonically (to electrons and/or muons).
\end{itemize}
In order to carry out realistic numerical simulations, we have used a 
modification of the MSSM implementation
\cite{SUSYWIG} of the {\tt HERWIG} event generator \cite{HERWIG}, 
in conjunction with the {\tt GETJET} code \cite{GETJET} for calorimeter
emulation and jet reconstruction. An {\tt ISAWIG} format \cite{ISAWIG}
input file has been edited by hand to incorporate the Higgs boson mass
spectrum and decay rates as predicted for each of the NMSSM points 1 -- 6, 
while in the main {\tt HERWIG}
code (v6.4) the subroutine implementing the vector-vector fusion process
has been modified to account for the 
different Higgs-$VV$ vertices pertaining to
the NMSSM points 1 -- 6 given in table~\ref{tpoints}. 
The above codes do not include $K$ factors for either the signal
or the background.  

\begin{figure}[p]
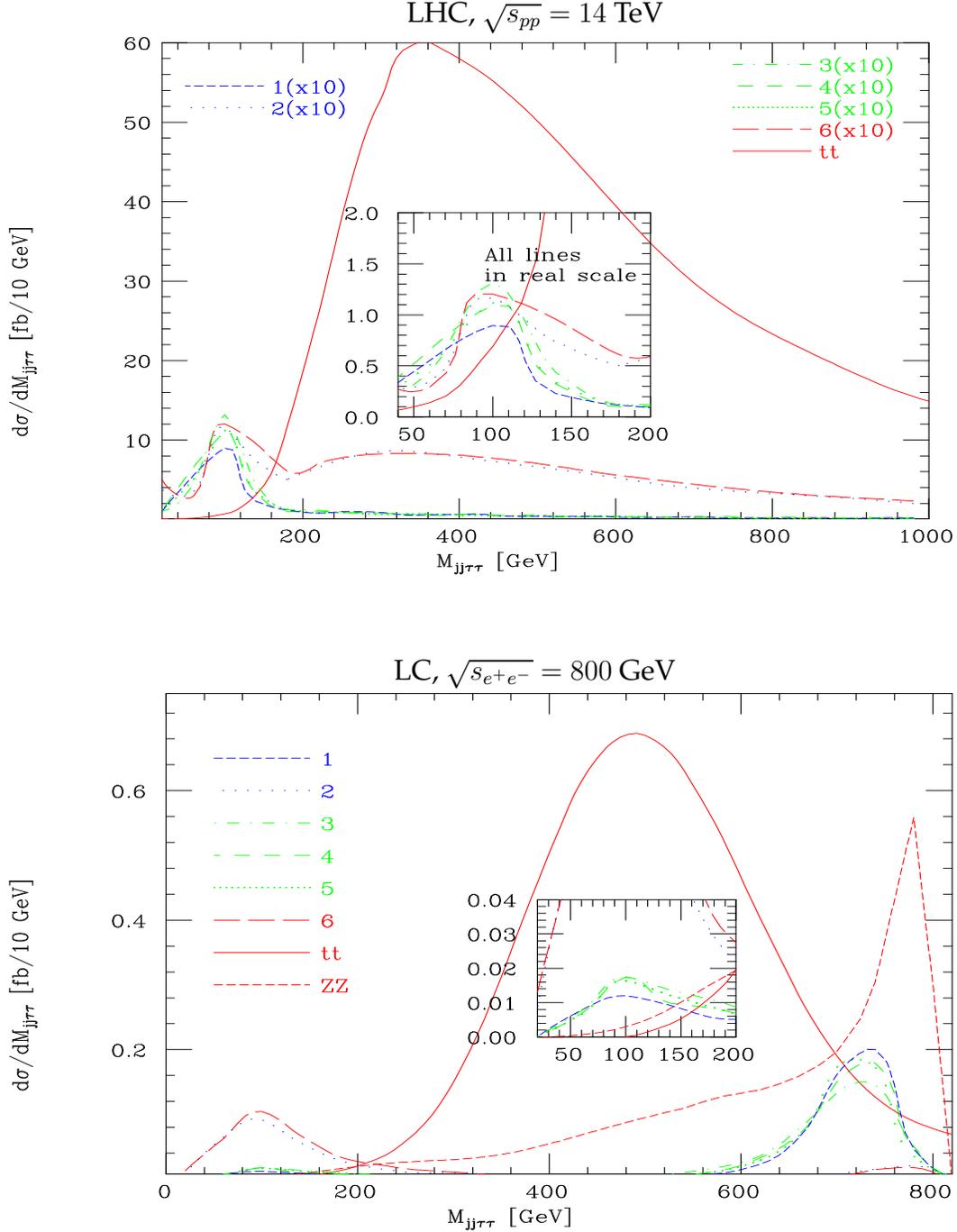

\begin{center}
\centerline{$~~~~~~~~~~~~~~~$LHC, $\sqrt{s_{pp}}=14$ TeV}
\centering\epsfig{file=AllMH_LHC.ps,angle=90,height=8cm,width=14cm}

\vspace*{1.0truecm}

\centerline{$~~~~~~~~~~~~~~~$LC, $\sqrt{s_{e^+e^-}}=800$~GeV}
\centering\epsfig{file=AllMH_LC.ps,angle=90,height=8cm,width=14cm}
\centerline{}

\noindent
\vspace{-1.0cm}
\caption{\footnotesize Reconstructed mass of the $jj\tau^+\tau^-$ system for signals and backgrounds after the selections described, at the LHC (top) and a LC (bottom). We plot $d\sigma/dM_{jj\tau^+\tau^-}$ [fb/10~GeV] vs $M_{jj\tau^+\tau^-}$~[GeV] using {\tt GETJET} and {\tt HERWIG 6.4} with:
{\tt IPROC=3720}
adapted to NMSSM couplings and decay rates for the signal and
{\tt IPROC=1706} for the background at the LHC; 
 {\tt IPROC=920}
adapted to NMSSM couplings and decay rates for the signal and
{\tt IPROC=126(250)} for the $t\bar t[ZZ]$ background at a LC). Statistics used: 500,000 points. Normalization is to the total cross section after cuts.
In both plots, the lines corresponding to points 4 and 5
are visually indistinguishable. No $K$ factors are included.}
\label{MH}
\end{center}
\end{figure}

An outline of the selection procedure and cuts used is the following.
\begin{itemize}
\item Acceptance cuts:
%\vspace{-0.25cm}
$$
|\eta_{\rm{jet}}|<5,
\quad
p^T_{\rm{jet}}>20~{\rm{GeV}},
\quad
\Delta R_{\rm{jet-jet}}>0.7,
$$
$$
\eta^{\rm{max}}_{\rm{jet}}\cdot\eta^{\rm{min}}_{\rm{jet}}<0,
\quad
|\eta_{\rm{lepton}}|<2.5,
$$
$$
p^T_{\rm{lepton}}>10~{\rm{GeV}},
\quad
{\rm{no~lepton~isolation}}.
$$
\item Since the $a_1$ will not 
have been detected previously, we must assume a value for $m_{a_1}$. It will be
necessary to repeat the analysis for densely spaced $m_{a_1}$
values and look for the $m_{a_1}$ choice that produces
the best signal. 

\item We look among the central jets for the combination with invariant mass
$M_{jj}$ closest to $m_{a_1}$ (no $b$-tagging is enforced, $b$'s
are identified as non-forward/backward jets). 

\item Select the two highest transverse-momentum leptons in any flavor
combination
and with opposite charge. After ensuring that these are not back-to-back
(by requiring that their relative angle is smaller than 175 degrees), 
resolve the $p^T_{\rm miss}$ along their directions and reconstruct
the invariant mass, $M_{\tau^+\tau^-}$. 

\item Plot the $M_{jj\tau^+\tau^-}$ invariant mass using the
four four-momenta reconstructed in the two previous steps, as seen
in the top plot of Fig.~\ref{MH} --- the plot presented assumes that
we have hit on the correct $m_{a_1}$ choice.

\end{itemize}

The selection strategy adopted is clearly efficient in reconstructing
the $h_1$ (for points 1--3) and $h_2$ (for points 4--6)
masses from the $jj \tau^+\tau^-$ system, as one
can appreciate by noting the peaks appearing in the LHC plot of
Fig.~\ref{MH} at $M_{jj\tau^+\tau^-}\approx100$~GeV. In contrast,
the heavy Higgs resonances at $m_{h_2}$ for points 1--3 and the rather light
resonances at $m_{h_1}$ for points 4--6 (recall table~\ref{tpoints}) do
not appear, the former mainly because of the very poor production rates and
the latter due to the fact that either the $h_1\to a_1 a_1$ decay mode
is not open (points 4, 5) or -- if it is -- the $b$-quarks and 
$e/\mu$-leptons eventually emerging
from the $a_1$ decays are too soft to pass the acceptance cuts
(point 6, for which $m_{a_1}=7$~GeV and $m_{h_1}=51$~GeV). 
For all six NMSSM setups, one
may see a hint of a resonance in the data in the very end of the
low mass tail of the $t\bar t$ background (see the insert in the top
frame of Fig.~\ref{MH}).
However, after summing the background distribution and any one of the
signal spectra, it could be difficult to ascertain the 
existence of the $h_1$ or $h_2$ peaks from the net line shapes.
Still, statistics are significant. 
To estimate $S/\sqrt B$, we assume $L=300~{\rm fb}^{-1}$,
a $K$ factor of 1.1 for $WW$ fusion
and a $K$ factor of 1.6 for the $t\anti t$ background.
(These $K$ factors are not included in the plots of Fig.~\ref{MH}.)
We sum events over the region $60\leq M_{jj\tau^+\tau^-}\leq 90$~GeV.
For points 1 -- 6 we obtain signal rates of about $S=890$, $600$, $750$,
$1030$, $915$, $500$, respectively.  The $t\anti t$ background
rate is $B\sim 320$. This gives
$N_{SD}=S/\sqrt B$ of 50, 34, 42, 58, 51, 28 for points 1 -- 6,
respectively. These are substantial.
However, given the broad distribution of
the signal, it is clear that the crucial question will be the accuracy
with which the background shape can be predicted from theory.  
The background {\it normalization} after the cuts imposed in our analysis
would be very well known from the higher $M_{jj\tau^+\tau^-}$ regions.

\subsection{The LC scenario}

While further examination of and refinements in the LHC analysis may ultimately
lead us to have full confidence in the viability of 
the NMSSM Higgs boson signals discussed above, 
an enhancement at low $M_{jj\tau^+\tau^-}$ of the type
shown (for some choice of $m_{a_1}$) will nonetheless be the only evidence
on which a claim of LHC observation of Higgs bosons can be based.
Ultimately, a means of confirmation and further study will be critical.
Thus, it is important to summarize
the prospects at the LC, with energy up to 800~GeV, 
for detecting the NMSSM Higgs bosons, especially in
the context of the difficult scenarios 1 --- 6 of table~\ref{tpoints}
discussed here.  As we summarize below, the LC is certainly guaranteed
to find the $h_1$ (points 1--3) or the $h_2$ (points 4--6) 
and will allow a good determination of the
$h_1a_1a_1$  or $h_2a_1a_1$ branching ratio, 
respectively, a possibly very important piece
of information for unraveling the model. In these scenarios, the LC
would be an absolutely essential complement to the LHC,
and, in particular, would be the only machine at which precision
Higgs physics studies could be pursued.
In what follows, we will use the notation $h$ 
for $h=h_1$ for points 1--3 and $h=h_2$ for points 4--6 in
table~\ref{tpoints}.

Because the $ZZh$ coupling is nearly full strength in all cases, 
and because the $h$ mass is of order 100~GeV,
discovery of the $h$ will be very straightforward via $e^+e^-\to Z h$ 
using the $e^+e^-\to ZX$ reconstructed $M_X$ technique which
is independent of the ``unexpected'' complexity of the $h$ decay
to $a_1a_1$. This will immediately provide a direct measurement
of the $ZZh$ coupling with very small error~\cite{review}. 
The next stage will be
to look at rates for the various $h$ decay final states, $F$,
and extract $BR(h\to F)=\sigma(e^+e^-\to Zh\to ZF)/\sigma(e^+e^-\to Zh)$.
For the NMSSM points considered here, the main channels would
be $F=b\anti b b\anti b$, $F=b\anti b \tau^+\tau^-$ and
$F=\tau^+\tau^-\tau^+\tau^-$.\footnote{Here, and in the following
discussion, $b\anti b$ should be replaced by $jj$
for points 2 and 6, where $j$ refers to any possible non-$b$
jet.}  At the LC, a fairly accurate
determination of $BR(h\to F)$ should be possible in all three cases.
This is important if we are to be able to determine
$BR(h\to a_1 a_1)$ independently.  The procedure is the following. From
\bea
BR(h\to b\anti b b\anti b)&=& BR(h\to a_1a_1)[BR(a_1\to b\anti b)]^2\nn\\
BR(h\to \tau^+\tau^-\tau^+\tau^-)&=&BR(h\to a_1a_1)[BR(a_1\to \tau^+\tau^-)]^2\nn\\
BR(h\to b\anti b\tau^+\tau^-)&=&2BR(h\to a_1a_1)BR(a_1\to b\anti b)BR(a_1\to \tau^+\tau^-)\,,
\eea
we see that $BR(a_1\to \tau^+\tau^-)/BR(a_1\to  b\anti b)$
can be determined using two independent ratios.  Assuming that
these are the only two decay channels (as can be checked
by looking for other final states), the requirement
$BR(a_1\to \tau^+\tau^-)+BR(a_1\to b\anti b)=1$
then allows extraction of the individual $BR$'s.  Once these
are known, we can use the above equations to determine $BR(h\to a_1a_1)$.
Certainly, this procedure will be carried out in the $e^+e^-\to Zh$
production mode.  However, we have not performed the required simulation
for this paper to determine the actual accuracy that can be achieved.

Instead, we have considered the equally (or perhaps more) useful
vector-vector fusion mode that will be active at a LC. In fact, 
at 800~GeV or above, it is the dominant Higgs boson production channel for
CP-even Higgs bosons in the intermediate mass range. Contrary to the
case of the LHC though, the dominant contribution (from $WW$
fusion) does not allow for forward and backward particle tagging,
as the incoming electron and positron convert into (anti)neutrinos,
which escape detection. Although the $ZZ$ fusion contribution 
would allow tagging of forward/backward $e^-$ and $e^+$,
the cross section is a factor of 10 smaller~(see Fig. 4 of Ref.~\cite{review}) 
in comparison. Still, since the $t\anti t$
background would not typically be accompanied by forward/backward
$e^+$ and $e^-$ in $e^+e^-$ collisions, unless initiated
by $\gam\gam$ interactions (which tend to produce
less energetic $t\anti t$ pairs), tagging of the  electron and positron
in the final state might offer a handle against the $t\bar t$ noise.
But, in any case it seems unnecessary to focus on $ZZ$ fusion
given that $t\anti t$ production at the LC
proceeds through EW interactions, 
rather than via QCD as at the LHC. Indeed, 
in going from the LHC to a LC, the $t\anti t$
production rate should be reduced (apart from parton distribution functions)
by a factor of ${\cal O}(\alpha_{\rm{s}}/\alpha_{\rm{em}})^2$ 
with respect to the vector boson fusion Higgs boson signal (which is an 
electroweak process at both the LC and the LHC).
A new background of importance does emerge at the LC, namely 
$ZZ$-pair production with one $Z$ decaying to $jj$
and the other to $\tau^+\tau^-$ pairs. The $ZZ$ background 
is negligible at the LHC mainly because associated
energetic forward/backward jets from QCD initial state radiation
are infrequent. At the LC, the $ZZ$ background plays
a more significant role and has been simulated in our {\tt HERWIG}
and (LC-adjusted) {\tt GETJET} numerical analysis.

At a LC, the optimal signature will thus be different than at the LHC
and a different set of selection criteria are needed. We choose
selection criteria that will retain both the $WW$ and the $ZZ$
fusion Higgs boson production processes.
\begin{itemize}
\item We require an 
even number of oppositely-charged leptons ($n_\ell=2,4...$). 
\item If $n_\ell\ge 4$ then we demand that two of these
must be an electron and a positron, so that the final state
would be consistent with having been
generated by forward/back\-ward $e^\pm$ tagging in $ZZ$-fusion and
$\tau^+\tau^-$ decays in which both $\tau$'s decay 
leptonically (to electron and muons).
\item
Finally, we require at least 2 central jets and significant $p^T_{\rm{miss}}$.
\end{itemize}
Given the required final state as above, we employ the following
selection procedure and cuts.
\begin{itemize}
\item Acceptance cuts:
%\vspace{-0.25cm}
$$
|\cos\theta_{\rm{jet}}|<0.990,
\quad
p^T_{\rm{jet}}>5~{\rm{GeV}},
\quad
\Delta R_{\rm{jet-jet}}>0.4,
$$
$$
\eta^{\rm{max}}_{e^+}\cdot\eta^{\rm{min}}_{e^-}<0
~({\rm{if}}~n_\ell\ge4),
\quad
|\cos\theta_{\rm{lepton}}|<0.995,
$$
$$
p^T_{\rm{lepton}}>5~{\rm{GeV}},
\quad
{\rm{no~lepton~isolation}}.
$$

\item We look among the central jets for the combination with invariant mass
$M_{jj}$ 
closest to $m_{a_1}$ (again, no $b$-tagging is enforced --- $b$'s
are identified as non-forward/backward jets).

\item Out of the $n_\ell$ leptons, upon excluding the $e^+e^-$
pair in which the electron and positron are those with the largest
rapidities if $n_\ell\ge4$, select the two with highest transverse-momenta 
in any flavor combination and with opposite charge. 

After ensuring that these are not back-to-back, 
resolve the $p^T_{\rm miss}$ along their directions and reconstruct
the invariant mass $M_{\tau^+\tau^-}$. 

\item Plot the $M_{jj\tau^+\tau^-}$ invariant mass (see the
bottom of Fig.~\ref{MH})
using the four four-momenta reconstructed in the two previous steps.

\end{itemize}
Note that it is not fruitful to place cuts on the invariant masses
$M_{jj}$ and $M_{\tau^+\tau^-}$ that exclude
$M_{jj},M_{\tau^+\tau^-}\sim m_Z$ in an attempt to reduce the $ZZ$
background. This is because the SM-like $h$ mass is 
typically of order $115$~GeV, \ie\ not so far from $m_Z$,
and the experimental resolutions in the two masses $M_{jj}$
and $M_{\tau^+\tau^-}$ are poor, either because of the large number
of hadronic tracks or the missing longitudinal momenta of the (anti)
neutrinos, respectively.

Just as at the LHC,
the peaks corresponding to the Higgs boson with weak $VV$
coupling (${h_2}$ for points 1 -- 3 and 
${h_1}$ for points 4 -- 6) do not appear in the $M_{jj\tau^+\tau^-}$
distribution which can be reconstructed at a LC. Here, though, for the
visible $h$ ($h_1$ for points 1 -- 3, $h_2$ for points 4 -- 6)
the situation is better than at the LHC.
For all points 1 -- 6 
the mass peaks (again centered at 100~GeV) are now very clearly visible
above both the $t\anti t$ and $ZZ$ backgrounds, particularly for the case of
points 2 and 6 (see insert in the bottom
frame of Fig.~\ref{MH}). 
Assuming $L=500~{\rm fb}^{-1}$, the points
1,3,4,5  yield 5 events per 10~GeV bin, on average, 
in the 50 to 150~GeV mass interval of interest. This would
constitute a convincing signal given the very small size predicted for the
background.
Notice that, although to assign the entire
missing transverse momentum to the $\tau$-lepton system may seem
not entirely appropriate
(as it is largely due to forward/backward 
(anti)neutrinos from the incoming electrons and positrons
in  $WW$ fusion),
this does not hamper the ability to reconstruct the Higgs mass peaks.
However, this implies that a proportion of the signal events will
tend to reproduce the overall $\sqrt{ s_{e^+e^-}}$
value in the $M_{jj\tau^+\tau^-}$ distribution. The effect
is more pronounced for points 1 and 3--5, which is where the 
$a_1$ mass is larger 
(see Table~\ref{tpoints}) so that most of the hadronic
tracks composing the emerging jets easily enter the
detector region. For points 2 and 6, where $m_{a_1}$
is below 10~GeV, this may often not be true and it appears that
the consequent effect of these hadrons escaping detection 
 is that of counterbalancing the $p^T_{\rm{miss}}$
contributions related to the neutrinos left behind in $WW$ fusion reactions.
For the case of the $ZZ$ noise,
in the presence of full coverage and perfect resolution of the detector,
one would have $M_{jj\tau^+\tau^-}\equiv\sqrt{s_{e^+e^-}}$, which
explains the very noticeable preference for this process 
to produce a concentration of events with $M_{jj\tau^+\tau^-}$
around 800~GeV in the mass distribution. (The ``tails'' beyond 
$\sqrt{s_{e^+e^-}}$ are due to the smearing of the visible
tracks in our Monte Carlo analysis.)

Finally notice that we have included Initial State Radiation (ISR)
and beam-strahlung effects, as predicted using the {\tt HERWIG} default.
These tend to introduce an additional unresolvable 
missing longitudinal momentum, although to a much smaller extent than 
do the Parton Distribution Functions (PDFs) in hadron-hadron scattering
at the LHC.

A final note regarding $WW,ZZ$ fusion to $h$.  Since we will
know from $Zh$ production the magnitude of the $WWh$ and
$ZZh$ coupling, the rate for $e^+e^-\to ZZ,WW\to h\to b\anti b\tau^+\tau^-$
will determine $BR(h\to b\anti b\tau^+\tau^-)$, with similar
results perhaps possible for other channels.  Thus, we could
hope to implement the procedure described earlier for extracting
$BR(h\to a_1a_1)$ using both $WW,ZZ\to h$ fusion and $Zh$
production.

\section{Conclusions}\label{sec:Conc}

In summary, if the NMSSM 
parameters are such that the most SM-like of the CP-even
Higgs bosons, $h$, is relatively light and decays primarily to
a pair of CP-odd Higgs states, $h\to aa$, then there
will be a statistically highly significant LHC signal
(from $WW\to h\to aa$) of an $S/B\sim (500-1000)/300$ bump 
(for $L=300~{\rm fb}^{-1}$) in the low-mass tail of
a rapidly falling $jj\tau^+\tau^-$ mass distribution.
The LHC would thus give
a very strong indication of the presence of a Higgs boson. However,
this detection mode is not exactly in the gold-plated category. 
The LC will be absolutely essential
in order to confirm that the enhancement seen at the LHC
really does correspond to a Higgs boson. At the LC, 
discovery of a light SM-like $h$ is guaranteed
to be possible in the $Zh$ final state using the recoil mass technique.
Further, we have seen that $WW,ZZ$ fusion production
of the $h$ will also produce a viable signal in the $jj \tau^+\tau^-$
final state (and perhaps in the $4j$ and $\tau^+\tau^-
\tau^+\tau^-$ final states as well, although we have not examined 
these~\cite{preparation}).\footnote{For many, but not all, parameter choices
of interest, $jj=b\anti b$.}
Measurement of the relative rates for the $4j$, $2j2\tau$ and $4\tau$
final states will allow a determination of the $a\to jj$
and $a\to \tau^+\tau^-$ branching ratios, which in turn will
make possible the determination of $BR(h\to aa)$, a potentially
very important measurement.  Unfortunately, the standard techniques
for determining the total width of a SM-like $h$ relying
on the $\gam\gam$ and/or $WW$ final state decays of the $h$
will not be available and it is thus unlikely that we can
convert a measurement of $BR(h\to aa)$ into a determination
of the partial width $\Gamma(h\to aa)$ (which would be most
directly related to the very interesting $haa$ coupling strength).
As we have stressed, for parameter space points of the type we
have discussed here, detection of any of the other MSSM
Higgs bosons is likely to be impossible at the LHC and
is likely to require an LC with $\sqrt{s_{e^+e^-}}$ above the relevant
thresholds for $h'a'$ production, where $h'$ and $a'$ are 
heavy CP-even and CP-odd Higgs bosons, respectively.
Although results for the LHC indicate that Higgs boson discovery
will be challenging for the type of situations we have considered,
improved techniques for extracting
a signal are likely to be developed once data is in hand and the $t\anti t$
background can be more completely modeled.  Clearly, if SUSY
is discovered and no Higgs bosons are detected in the standard
MSSM modes, a careful search for the signal we have considered
should have a high priority.
Finally, we should remark that the $h\to aa$ search channel
considered here in the NMSSM framework is also highly
relevant for a general two-Higgs-doublet model, 2HDM.  
It is really quite possible
that the most SM-like CP-even Higgs boson of a 2HDM
will decay primarily to two CP-odd states.  This
is possible even if the CP-even state is quite heavy, unlike
the NMSSM cases considered here.

\subsection*{Acknowledgments}
JFG is supported by the U.S. Department of Energy and the Davis
Institute for High Energy Physics. SM thanks the UK-PPARC for financial support
and D.J. Miller for useful conversations. 
CH is supported by the European Commission RTN grant HPRN-CT-2000-00148.
JFG, CH, and UE thank
the France-Berkeley fund for partial support of this research.

\end{document}